# Sinusoidal shaped hollow fibers for enhanced mass transfer

T. Luelf, M. Tepper, H. Breisig, M. Wessling

*RWTH Aachen University, Chemical Process Engineering, Turmstr. 46, 52064 Aachen, Germany*

*DWI Interactive Materials Research, Forckenbeckstrasse 50 52074 Aachen, Germany*

**Abstract**

Inducing secondary flows and vortices is known to enhance mass transport. They can be imposed by structured flow channels for instance. In particular, these vortices reduce fouling and concentration polarization. In this work we present a new method of producing hollow fiber membranes with a sinusoidal change in diameter over the fiber length. We engineered a pulsation module that imposes a sinusoidally fluctuating bore liquid flow rate. Harmonic bore flow conditions can be varied over a wide range. The fluctuating bore liquid flow rate translates into axial membrane properties varying with respect to inner bore diameter and wall thickness. The resulting narrowing and widening of the membrane lumen channel are hypothesized to induce secondary vortices to the liquid inside the membrane lumen known as the Bellhouse effect. Improved oxygen transport from shell-to-lumen side prove superiority over straight hollow fiber membranes in G/L absorption process. We anticipate the dynamic flow module to be easily integrated into currently existing hollow fiber membrane spinning processes.

**Todo list**

## 1. Background

### 1.1. Importance of hollow fiber geometry

Hollow fiber membranes offer important advantages over other membrane configurations. They show well-defined flow conditions on the inside of the fiber. Packing densities are superior over flat sheet based membrane modules. In fact, applications requiring mass production of membranes mainly rely on hollow fibers as their production and assembly into modules is highly scalable. This is particularly true for medical applications such as hemodialysis and blood oxygenation [1]. While commonly assumed to be done only by spiral-wound modules [2], even seawater desalination is done with hollow fiber membranes [3] at a scale as important as spiral-wound modules. Parallelization and high degrees of production automation allows for scalable spinning processes [4].

### 1.2. Hollow fiber spinning

In industrial hollow fiber fabrication the fiber is extruded through a spinneret [5]. The extrusion of the inner lumen flow channel is done parallel to the main flow direction of the polymer solution by engineering parallel flow inside the spinneret.

Although radial stress may occur due to die swell, the resulting geometry is a constant lumen channel cross-section along the main direction of the fiber. This indisturbed tubular geometry causes close to perfect laminar flow profiles in later applications. However, in applications like hemodialysis, the fibers are slighly undulated to maintain a good dialysate flow on the fiber shell side [6], while lumen side flow is assumed to remain undisturbed laminar. In membrane contactor modules, shell side baffles and other means are integrated into the hollow fiber module to allow for fluid flow perpendicular to the main fiber direction [7]. This allows for increased mixing on the shell side of the fiber only. No increased mixing is possible with this method on the fiber inside.

### 1.3. Mass transport in furrowed channels

The fluid flow in furrowed or grooved flow channels has been reported in literature since the work of Bellhouse [8] in 1973. These studies revealed that periodic geometrical flow disturbances can lead to increased mixing inside the flow channels. This can be transferred into an increase in transfer rate (either heat or mass transfer) across the flow channels surface [9, 10]. Early numerical and experimental studies agreed well [11, 12].

Recently, Kasisteropoulou [13] investigated the fluid flow in periodically grooved microchannels. Nishimura et al. investigated the fluid flow in wavy channel in numerous publications as early as 1991. They used both experimental [14–20] and numerical methods [14, 16–18, 20, 21] to investigate the flow behavior in wavy flat channels [14–18, 21] and wavy walled tubes [19, 20]. In both cases the formation of vortices have been reported. When comparing both geometries they conclude, that the mass transfer increment in the studied flow regime ($50 < Re < 1000$) is larger for the wavy-walled tube than the wavy-walled flat channel [19]. Neusser and coworkers recently attempted to transfer the concept of furrowed flow channels to new biohybrid membranes. They developed a setup to transfer a flat sheet oxygenation membrane into a membrane having furrows and seeded them with endothelial cells [22, 23].

Unfortunately, all these principles have not entered the fabrication process of hollow fiber preparation yet. It is the aim of this work to establish evidence that continuous preparation of hollow fibers having furrowed lumen channels is possible and offers the anticipated improvements in mass transport.

## 2. Materials and methods

### 2.1. Transient flow conditions in membrane spinning

To allow for a membrane geometry that combines static mixing by furrowed channel geometry with a hollow fiber membrane fabrication process, we choose to design a sinusoidal variation in fiber lumen diameter along the fiber length. The variation is realized by adding a fluctuation in flow to a constant bore fluid flow rate as shown in Figure 1. The constant flow of the polymer solution was induced by a HARVARD syringe pump. Pulsation was induced by a custom made pulsation module shown in Figure 2: through a T-piece, liquid volume is pressed and withdrawn alternatively into and from the lumen pipe to reach the desired pulsation. This volume was pumped using a microliter syringe incorporated into the pulsation module. Using the variable rotational speed of an electrical stepper motor, the rotation is transferred into a linear motion by a linear slider attached to a disk with multiple eccentric drillings. The linear slider allows for the attachment of the syringe plunger. Now the frequency of fluid flow can be adjusted by the rotation speed of motor while the amplitude can be varied by the choice of eccentricity on the turning plate. The amplitude of pulsation can also be varied by the selection of the microliter glass syringe (Hamilton), thus pulsation volume can be selected from the whole spectrum of microliter syringes available.

Due to the mechanical design, the bore pulsation was sinusoidal, as can be derived by transformation of rotational movement with constant frequency towards linear movement. The superposition of a constant bore flow rate with a sinusoidal pulsation allows for a decoupled net flow rate from pulsation frequency and amplitude, thus offering all desired degrees of freedom for the pulsating flow manipulation, namely:

- Amplitude $a$ of pulsation
- Frequency $f$ of pulsation
- Net bore flow rate

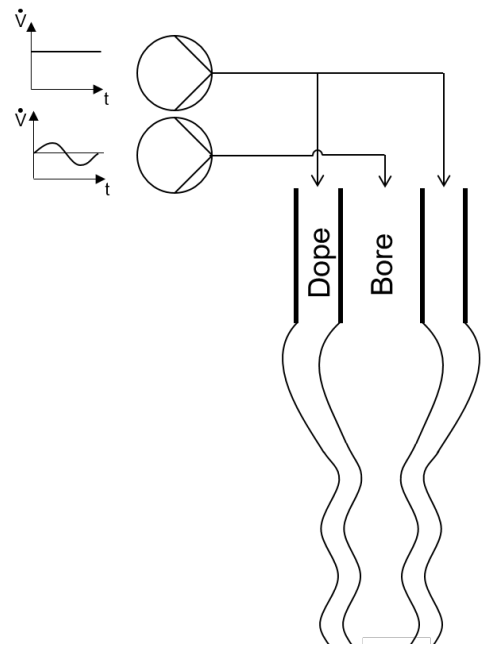

Figure 1: Pulsation concept for the production of sine shaped hollow fibers. The flow rate of polymer solution is constant whereas bore liquid flow rate is a function of time, leading to variable axial fiber diameters

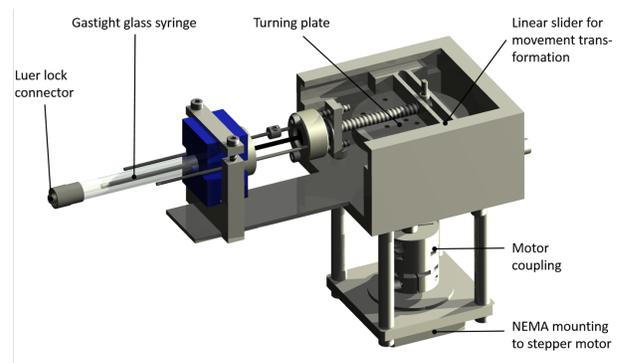

Figure 2: Custom made pulsation module consisting of: 1: Luer Lock connector for plug and play use in state-of-the-art spining lines, 2: HAMILTON gastight glass syringe, 3: turning plate with discrete eccentric drillings for amplitude manipulation, 4: linear slider for transformation to linear movement, 5: Motor coupling, 6: Mounting plate for NEMA connection to stepper motor. Stepper motor not shown.

| Solution | PVDF [wt.%] | LiCl [wt.%] | NMP [wt.%] | Ethanol [wt.%] | H$_2$O [wt.%] |
|----------|-------------|-------------|------------|----------------|---------------|
| Polymer  | 17          | 4           | 79         | -              | -             |
| Bore     | -           | -           | -          | 50             | 50            |

Table 1: Composition of polymer and bore solution applied in the spinning experiments

| Sample | $V_{Bore}$ [mL/min] | $V_{Polymer}$ [mL/min] | $c_{Wheel}$ [cm/s] |
|--------|---------------------|------------------------|--------------------|
| 1-8    | 0.6                 | 0.9                    | 2                  |
| 9-10   | 0.6                 | 1.35                   | 2.33               |

Table 2: Conventional spinning parameters used in this work

| $d_{Bore}$ [mm] | $d_{pol-in}$ [mm] | $d_{pol-out}$ [mm] |
|-----------------|-------------------|--------------------|
| 0.51            | 0.82              | 1.12               |

Table 3: Spinneret dimension

## 2.2. Pulsation calculation

Assuming ideal incompressible fluid behaviour of the bore fluid, a superposition of two flows by a T-piece results in a mathematical superposition of flow rates. The flow rate can be quanitified by equation 1:

$$\dot{V}_{bore,out}(t) = \dot{V}_{bore,conti} + \dot{V}_{bore,pulse}(t). \tag{1}$$

The time dependant pulsating contribution is defined by the radius $r_{pulse}$ set at the turning plate and the syringe plunger diameter in combination with the constant rotational speed. Rewriting equation 1 results in:

$$\dot{V}_{bore,pulse} = A_{pulse} \cdot c_{pulse} = A_{pulse} \cdot 2\pi n \cdot r_{pulse} \cdot cos(2\pi n \cdot t), \tag{2}$$

Integrating the ratio of bore and polymer solution flow rate and the fiber drawing velocity an ideal spinning process with no change in dimensions over the air gap distance would result in fiber dimensions as follows:

$$d_{hf,i} = \sqrt{4 \cdot \frac{\dot{V}_{Bore}}{\pi \cdot c_{hf}}} \tag{3}$$

$$d_{hf,o} = 2 \cdot \sqrt{\frac{\dot{V}_{Dope}}{c_{hf} \cdot \pi} + \frac{d_{hf,i}^2}{4}}, \tag{4}$$

with $d_{hf,i}$ being the inner fiber diameter and $d_{hf,o}$ being the outer diameter. The pulsating membrane segments are defined by each full sinusoidal run. The period of the sine wave transfers to a length of the wave in the final membrane diameter.

$$l = \frac{1}{2} \cdot c_{hf} \cdot T = \frac{1}{2} \cdot c_{hf} \cdot n^{-1}, \tag{5}$$

with $T = n^{-1}$.

## 2.3. Membrane fabrication

Hollow fiber membranes have been produced on a single fiber spinning line. Polymer solutions have been prepared on the basis of Solvay Solef 6010 PVDF according to Table 1. LiCl (Cas.-Nr. 7447-41-8) was supplied by Carl Roth, NMP (Cas.-Nr. 872-50-4) was supplied by Fisher Scientific.

The spinning line was adapted to pulsating spinning by adding the pulsation module described above without further modifications. In order to maintain comparability, hollow fibers have been spun with active pulsation and without pulsation. Spinning conditions are listed in Table 2. A Hamilton gas tight syringe with 250µl holdup and plunger diameter of 2.3 mm was used on the pulsation module.

## 2.4. Image analysis

Membrane geometries have been measured with an optical device specifically designed for determining fiber dimensions. The digital information was further analyzed via MATLAB® image analysis if not stated otherwise. Fiber walls were made transparent by soaking with octanol, to introduce a fluid of similar refractive index into the pores of the membrane matrix. After soaking the fiber became transparent [24]. H$_2$O colored with red dye was added into the fiber lumen to create contrast between the lumen and membrane matrix. In order to prevent refractive distortion, the photographic image was taken with the fiber submerged in a petri-dish filled with octanol, thus providing a planar liquid surface.

## 2.5. Gas-liquid contacting

To evaluate the potential of mass transfer enhancement inside the hollow fiber lumen, mass transfer experiments have been performed. Gas absorption into a flowing liquid has been chosen as experimental system. Experiments have been performed via oxygen-water gas-liquid contacting. The experiments were performed in a setup as illustrated in Figure 3.

Water is pumped from a glass vessel through the metal tubing through the membrane module and back into the glass vessel. Oxygen atmosphere was applied on the shell side of the module at ambient pressure. The oxygen level inside the vessel was continuously measured via a Pyroscience Robust Probe OXROB10. Between each measurement run the vessel has been degassed by nitrogen both inside the vessel and on the outside of the membrane module. Control measurements to ensure no oxygen transfer through tubing and sealing have been performed regularly with a metal tube instead of the membrane module.

| Membrane ID | $n$ [Hz] | $r_{puls}$ [mm] |
|-------------|----------|-----------------|
| 1           | 1.17     | 7.5             |
| 2           | 0        | -               |
| 3           | 0.78     | 7.5             |
| 4           | 0        | -               |
| 5           | 1.17     | 7.5             |
| 6           | 1.56     | 7.5             |
| 7           | 0.78     | 7.5             |
| 8           | 1.17     | 10              |
| 9           | 1.17     | 7.5             |
| 10          | 0        | -               |

Table 4: Frequency and amplitude bore flow conditions used to obtain different membrane geometries

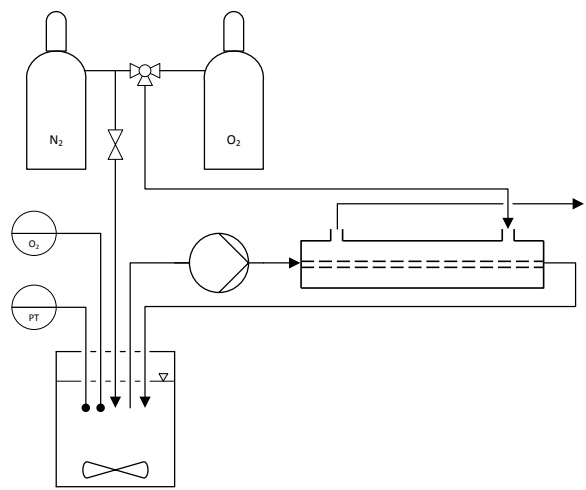

Figure 3: Flowpath of gas-liquid contacting experimental setup. The oxygen content in the stirred vessel is measured over time. Nitrogen or oxygen are connected to the membrane module as demanded. Specific flow rates are set with a controllable gear pump.

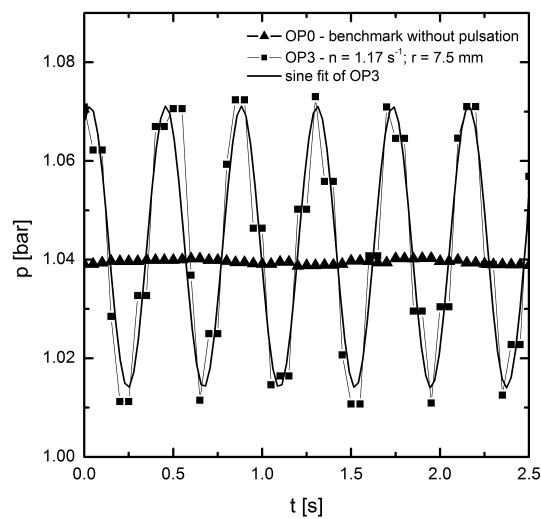

Figure 4: Bore fluid pressure as indication for sinusoidal flow rate with minimal damping effects at the spinneret. ▲ Measurement of constant flow rate operation; ■ Measurement of pulsating flow

## 3. Results

### 3.1. Pulsation evaluation

The behaviour of the pulsation module was tested during membrane preparation by pressure measurements in the bore channel of the spinneret assembly. As bore liquid is considered to be a Newtonian, incompressible fluid, a sinusoidal pulsation in flow rate should directly correlate to a sinusoidal pulsation in pressure, as given by Bernoulli's law.

The resulting pressure variation is shown in Figure 4. The data points represent actual measurements of pressure, whereas the lines are fitted sine functions. Double data points directly next to each other are caused by the restricted time resolution of the data acquisition device. Triangle points are measurement data from a standard spinning process with the same spinning conditions. Squared data points were obtained during pulsating operation. The pressure fluctuations correlate well with the analytical sine functions and only little to no damping is observed due to stiff tubing and incompressible medium and parts. At this point we conclude that the pulsation induction into the bore fluid line of the spinning setup is technically viable.

### 3.2. Membrane morphology

During the spinning process, it became immediately clear that the lumen volume dilation and contraction modified the geometry of the hollow fiber after coagulation. Fiber dimensions will be analyzed below in more detail. Figure 5 shows the cross section of Sample 5 at axial positions near maximum and minimum diameter. It is apparent that the fiber wall thickness is higher for the smaller inner diameter section. This redistribution of wall thickness is a direct result of a radial conservation of mass distributed over larger cross-sectional area of the fiber wall. Whether there is also a redistribution of mass in the axial wall direction cannot be concluded at this stage. The corrugated inner morphology was studied in detail by Bonyadi et al. [25]. Different conditions under which corrugated cross-sectional areas occur are suggested. Among other hypothesises possibly leading to a corrugated inner cross-section, the contribution of elastic and buckling instability is enhanced in our process. In addition to an inward radial force induced by a shrinkage of polymer dope, the fluctuating bore fluid adds a contribution to this force.

### 3.3. Membrane geometries

Figure 6 shows the fiber diameters gained by optical image analysis. Table 5 shows the results of the corresponding sine function fitting. The fiber diameters were fit to comply with equation 6. The periodical shape of the fibers is visible. Higher pulsation rates during membrane fabrication are represented by higher frequency of diameter-change in fiber measurements. Indications of some non-ideal sinusoidal shape in the samples of Figure 6 b) are visible. Also fibers with the same pulsation setting (Sample 3 and 7) in some cases show different amplitudes in the diameter measurement, resulting from unequal pulsating over the circumference of the fiber. These effects of non-ideal behavior require further deeper analysis and are beyond the scope of the current manuscript.

$$f(x) = a \cdot sin(2\pi \cdot f \cdot x) + d_m \qquad (6)$$

Figure 7 presents the inner fiber lumen of Sample 4, 6 and 9, spun with induced bore pulsation. A sinusoidal variation

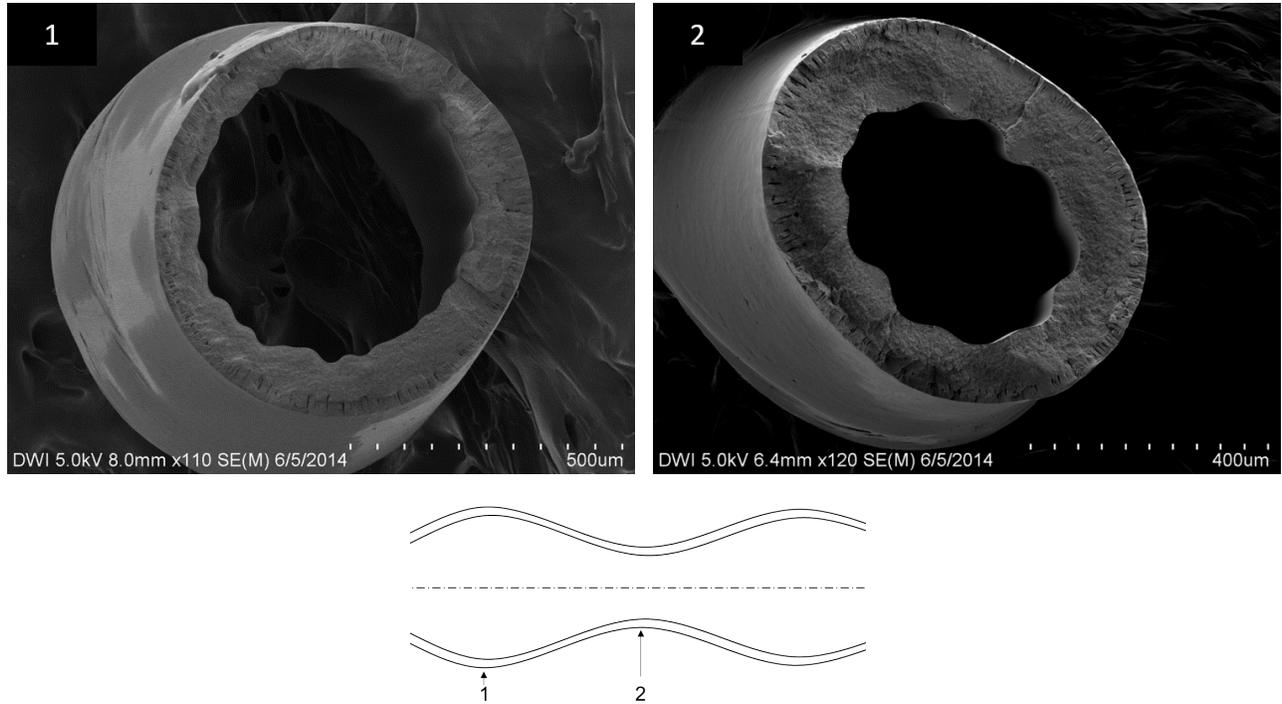

Figure 5: FESEM images of fiber sample 5 cross-sections taken on the thick (1) and thin (2) element of the fiber.

|   | $a \cdot sin(2\pi \cdot f \cdot x) + d_m$ | | |
|---|---|---|---|
| # | a [mm] | f [1/mm] | $d_m[mm]$ |
| 1 | 0.243 | 0.139 | 1.00 |
| 3 | 0.215 | 0.092 | 0.81 |
| 5 | 0.211 | 0.136 | 1.06 |
| 6 | 0.149 | 0.187 | 1.10 |
| 7 | 0.220 | 0.092 | 1.02 |
| 8 | 0.248 | 0.141 | 1.01 |
| 9 | 0.039 | 0.13 | 1.06 |

Table 5: Fitted parameters of fiber outer diameter calculated from image analysis

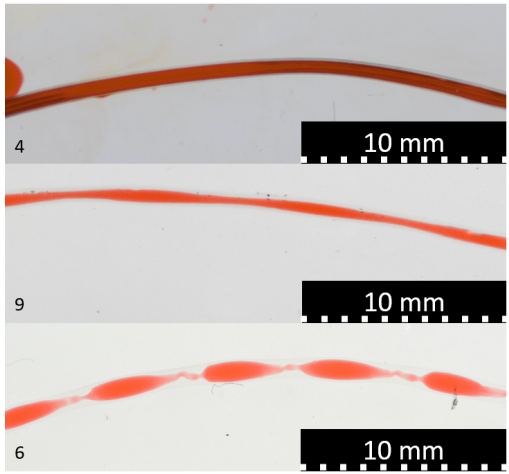

Figure 7: Fiber lumen of Sample 4, 9 and 6 spun with induced bore pulsation. Membrane wall is transparent due to refractive index matching of PVDF with octanole

in inner fiber diameter can be observed with an connect lumen channel between wide and narrow sections. The geometry is regular without visible distortions from sinusoidal diameter variation for fiber 9 with relatively low pulsation frequency of 1.17 Hz compared to 1.56 Hz of Sample 6. Sample 6 shows non-idealities of the bore channel of the final fiber. The small diameter regions are narrow flow channels with diameter changes that do not follow a sine function. As the pulsation rate is 1.56 Hz and thus significantly higher than for sample 9 the flow rate of the pulsation device leaving the bore fluid line overcompensated the constant flow rate of the syringe pump, leading to partial backflow through the nascent fiber into the spinneret maintaining an open channel with minimal diameter.

### 3.4. Mass transfer experiments

Gas-liquid contacting experiments have been performed with different fiber batches, both with straight and sine shaped geometry. Mass transfer rates as a function of lumen flow rates are plotted in Figure 8. Positive deviation of the transfer rates become clear for higher lumen flow rates. The calculated flow rates are total flow rates.

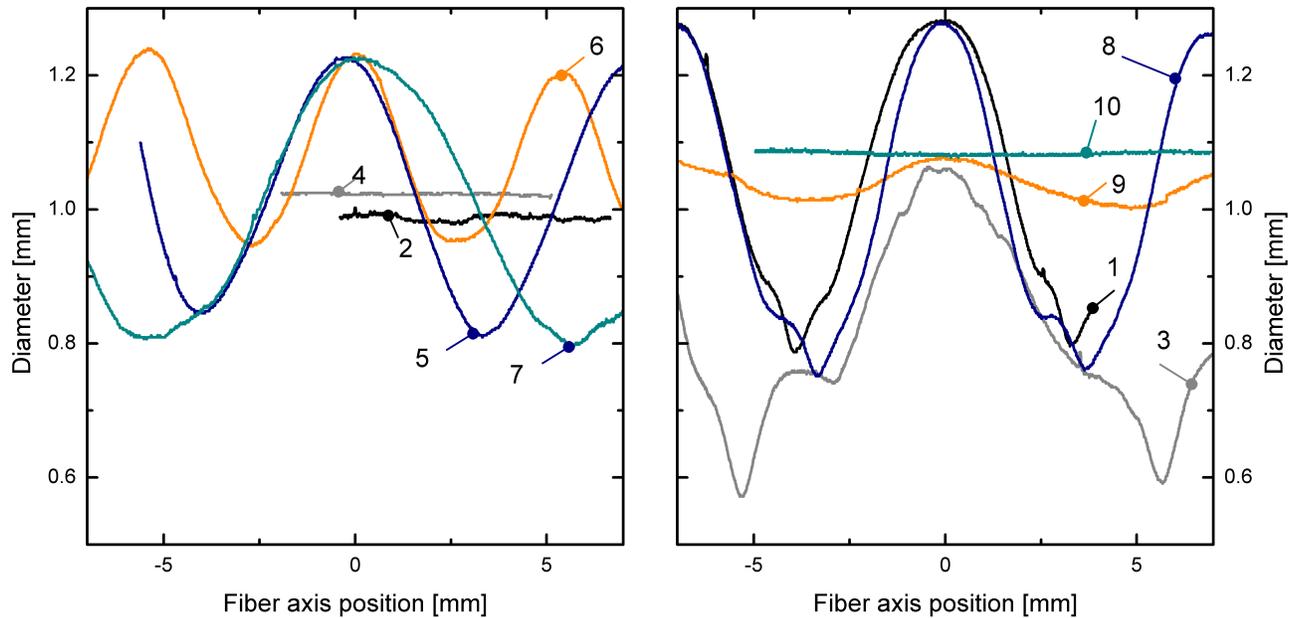

Figure 6: Plot of fiber outside diameters along the fiber axis. Diameter maximum values are set to be at zero axis position. Colours and numbers denote different samples.

For low flow rates all membrane samples have about the same oxygen flux up to a flow rate of approximately 2 mL/min. At this flow rates the transfer rate of the highly pulsating fiber 6 outperforms fibers 9 and 4 e.g. with oxygen fluxes of 8.1 mmol/m$^2$ h for fiber 6 and 5.4 mmol/m$^2$ h and 3.9 x10 mmol/m$^2$ h for fibers 9 and 4 respectively.

## 4. Discussion

A pulsation module has been designed to account for a highly scalable and flexible flow pulsation to be applicable as a plug-in solution on an existing membrane spinning line.
Pulsation of the bore fluid was successfully transferred to the precipitating polymer solution. The axially variable geometry was conserved into the final fiber. The created degrees of freedom in membrane spinning can be used in multiple ways, e.g. to create fibers with non sinusoidal diameter variation by changing the motor movement to a non constant rotation. While the results of mass transfer are promising, the current work can only be considered as a proof of principle. Many potential further steps can be taken based on the above data, but need to be evaluated in more detail.

- Transferring the principle to other material systems.
- Exploration of high frequency to find out how the frequency influences geometry and mass transport.

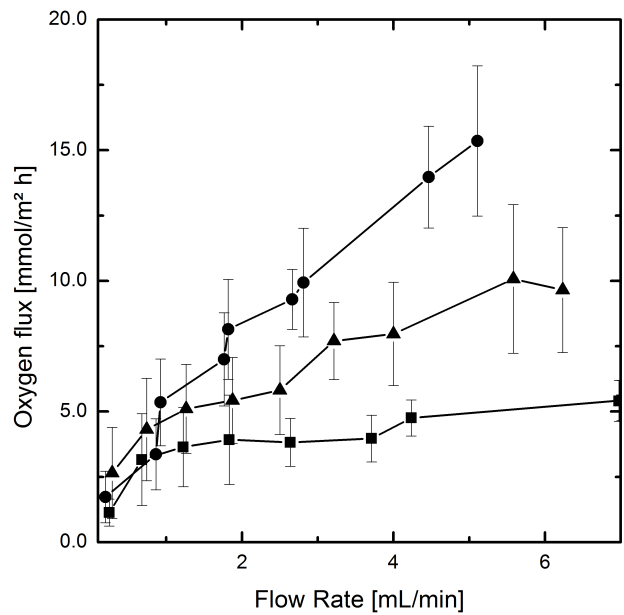

Figure 8: Results of oxygen transfer experiments. ● Membrane batch 6; ▲ Membrane batch 9; ■ Membrane batch 4. Error bars indicate standard deviation obtained by at least triplicate measurements.

- Applying the proposed fiber geometry in membrane filtration applications to influence polarisation phenomena.
- Transferring the concept to a multi-spinneret line where one plunger system actuates multiple spinnerets being operated out of one block as is found in industrial parallelized spinning processes.
- Establishing fluctuation conditions of the bore fluid which are asymmetric in nature and thus shape non-sinusoidal geometries.
- Interaction of pulsatile feed flow with the new structures: Does it improve mass transport even further by inducing non-steady eddies that interfere with the laminar stream lines and diffusion resistances.
- In the past we have reported that strong gradients in the skin of microstructured fibers need to be avoided [26] and only subtle transitions guarantee sharp separation characteristics. How does axial rather than radial gradients influence separation skin properties?

## 5. Conclusion and Outlook

The presented work describes the implementation of a novel method to produce hollow fiber membranes with passive mixing functionalities. The mixing is achieved via a sinusoidal variation in diameter along the fiber length. Gas liquid contacting with this novel hollow fiber membrane shows promising improvements. Transfer rates increased by a factor of about 2.4.

While all presented experiments have been performed with a sinusoidal plunger movement of the pulsation device, the movement can be altered in future work. A higher pulsating frequency with a smaller microliter syringe would lead to more open fibers with intensified flow disturbance. On the other hand a time gap between each sine run or a change in pulsating frequency (e.g. fast collapsing of fiber diameter and slow backpumping) might lead to even more improved fiber properties or fibers with non symmetric structures.

### Acknowledgements

M. W. acknowledges the support through an Alexander-von-Humboldt Professorship. This work was performed in part at the Center for Chemical Polymer Technology CPT, which is supported by the EU and the federal state of North Rhine-Westphalia (grant no. EFRE 30 00 883 02).